# Colossal dielectric constant up to GHz at room temperature

S. Krohns,[1] P. Lunkenheimer,[1*] Ch. Kant,[1] A. V. Pronin,[2] H. B. Brom,[3] A. A. Nugroho,[4] M. Diantoro,[5] A. Loidl[1]

[1]Experimental Physics V, Centre for Electronic Correlations and Magnetism, University of Augsburg, 86159 Augsburg, Germany
[2]Hochfeld-Magnetlabor Dresden (HLD), Forschungszentrum Dresden-Rossendorf (FZD), 01314 Dresden, Germany and A. M. Prokhorov Institute of General Physics, RAS, 119991 Moscow, Russia
[3]Kamerlingh Onnes Laboratory, Leiden University, P.O. Box 9504, 2300 RA Leiden, The Netherlands
[4]Department of Physics, Bandung Institute of Technology, Jl. Ganesha 10 Bandung 40132, Indonesia
[5]Department of Physics, State University of Malang, Jl. Surabaya 6 Malang, 65145, Indonesia

*To whom correspondence should be addressed. E-mail: peter.lunkenheimer@physik.uni-augsburg.de

The search for new materials with extremely high ("colossal") dielectric constants, required for future electronics, is one of the most active fields of modern materials science. However, the applicability of the colossal-$\varepsilon'$ materials, discovered so far, suffers from the fact that their dielectric constant, $\varepsilon'$, only is huge in a limited frequency range below about 1 MHz. In the present report, we show that the dielectric properties of $La_{15/8}Sr_{1/8}NiO_4$ surpass those of other materials. Especially, $\varepsilon'$ retains its colossal magnitude of >10000 well into the GHz range. This material is prone to charge order and this spontaneous ordering process of the electronic subsystem can be assumed to play an important role in the generation of the observed unusual dielectric properties.

In recent years, large effort has been put into the development and characterization of new colossal-$\varepsilon'$ materials. For example, the recent discovery (1,2) of "colossal" values of the dielectric constant, $\varepsilon'$, up to about $10^5$ in $CaCu_3Ti_4O_{12}$ (CCTO) has aroused tremendous interest and a huge number of publications deals with its investigation and optimization. Aside of the extensively investigated CCTO, there are also some reports of other colossal-$\varepsilon'$ materials (e.g., refs. (3,4,5,6,7)), mainly transition metal oxides. While there is no clear definition, the term "colossal" typically denotes values of $\varepsilon' > 10^4$. Such materials are very appealing for the further miniaturization of capacitive components in electronic devices and also in giant capacitors that may replace batteries for energy storage.

Of course, colossal dielectric constants are also found in ferroelectrics where close to the phase transition very large values are reached. However, ferroelectrics are characterized by a strong temperature dependence of $\varepsilon'$ around their critical temperature, which restricts their applicability. In contrast, CCTO and other materials stand out due to their colossal-$\varepsilon'$ values being nearly constant over a broad temperature range around room temperature. But in all these materials a strong frequency dependence is observed, revealing the signature of relaxational contributions, namely a steplike decrease of $\varepsilon'$ above a certain, temperature-dependent frequency, accompanied by a peak in the dielectric loss. Intrinsic relaxations are commonly observed, e.g., in materials containing dipolar molecules, which reorient in accord with the ac field at low frequencies, but cannot follow at high frequencies. However, the extensive investigations of CCTO, have quite clearly revealed that the observed relaxation features are due to a non-intrinsic effect, termed Maxwell-Wagner (MW) relaxation (8,9,10). It arises from heterogeneity of the sample, which is composed of a bulk region with relatively high conductivity and one or several relatively insulating thin layers. The equivalent circuit describing such a sample leads to a relaxation-like frequency and temperature dependence (10). The insulating layers can arise, for example, from surface effects (e.g., depletion regions of Schottky diodes at the electrodes) or internal barriers (e.g., grain boundaries). However, this is rather irrelevant from an application point of view (e.g., external surface layers are used to enhance the capacitance in ferroelectrics-based multi-layer ceramic capacitors). Thus, although in CCTO the exact mechanism is not yet finally clarified, the interest in this material is still high. This is, amongst others, demonstrated by the fact that since its discovery in 2000, twelve so-called "highly-cited" papers on this topic have appeared (source: ISI Web of Science, Nov. 2008). Unfortunately, at room temperature the relaxation in CCTO leads to a decrease of $\varepsilon'$ in the MHz region and around GHz only values of the order of 100 are observed (8,11,12). In contrast, electronic applications, e.g., in computer technology or telecommunications nowadays require much higher frequencies, up to the GHz range.

In a search for a better material, we have investigated various compounds and found that $La_{15/8}Sr_{1/8}NiO_4$ (LSNO) is much more promising for fulfilling the requirements of modern electronics than CCTO. Early investigations of the system $La_{2-x}Sr_xNiO_4$ were triggered by the discovery of high-$T_c$ superconductivity in the isostructural $La_{2-x}Sr_xCuO_4$ (see, e.g., ref. (13)). While not superconducting, $La_{2-x}Sr_xNiO_4$ attracted much interest due to the formation of charge order in large parts of its phase diagram (14,15,16). In the present work, we report the results of broadband dielectric measurements of single crystalline LSNO in a frequency range up to 1 GHz. The LSNO crystals were prepared by the travelling solvent floating zone growth technique (17). For the dielectric measurements, frequency-response analysis and a coaxial reflection technique were used (17,18).

Figure 1 shows the temperature dependence of the dielectric constant, conductivity ($\sigma$) and loss tangent (tan $\delta$) measured with silver-paint contacts at various frequencies. Fig. 1A reveals a strong step in $\varepsilon'(T)$ from low-temperature values of about 300 up to colossal values of about 17000. It shifts to higher temperatures with increasing frequency and at the point of inflection of $\varepsilon'(T)$, the conductivity and loss tangent (Figs. 1B and C, respectively) show well-pronounced peaks. These are the typical signatures of a relaxation and the overall behaviour qualitatively resembles that of CCTO as shown, for example, in ref. (2). However, already in this temperature-dependent plot, strong quantitative differences in the relaxation behaviour become evident: While in CCTO the relaxation step at, e.g., 1 MHz occurs at about 200 K, in LSNO it is observed at a much lower temperature of about 60 K and, thus, the region of constant $\varepsilon'$ is much broader. In addition to this main relaxation feature, a second smeared-out relaxation step is observed in $\varepsilon'(T)$ (Fig. 1A) leading to values at high temperatures and low frequencies of about 50000. Correspondingly, broad shoulders in $\sigma$ and tan $\delta$ show up. Interestingly, if CCTO is investigated in a sufficiently broad temperature and frequency range, also there a second relaxation is detected (8,11,12,19). It should be noted that the loss tangent of LSNO (also termed dissipation factor; Fig. 1C), which should be minimised in any capacitor material, is at high frequencies smaller than in CCTO (2). The inset of Fig. 1 shows the temperature-dependent dielectric constant of the same LSNO crystal, after removing the silver-paint contacts and applying sputtered gold contacts. Here, truly colossal values of $\varepsilon'$ are found, reaching nearly $10^6$. Also in CCTO a strong enhancement of $\varepsilon'$ for sputtered contacts was observed and taken as indication of a surface-related origin of the colossal magnitude of the dielectric constant (8,11,12).

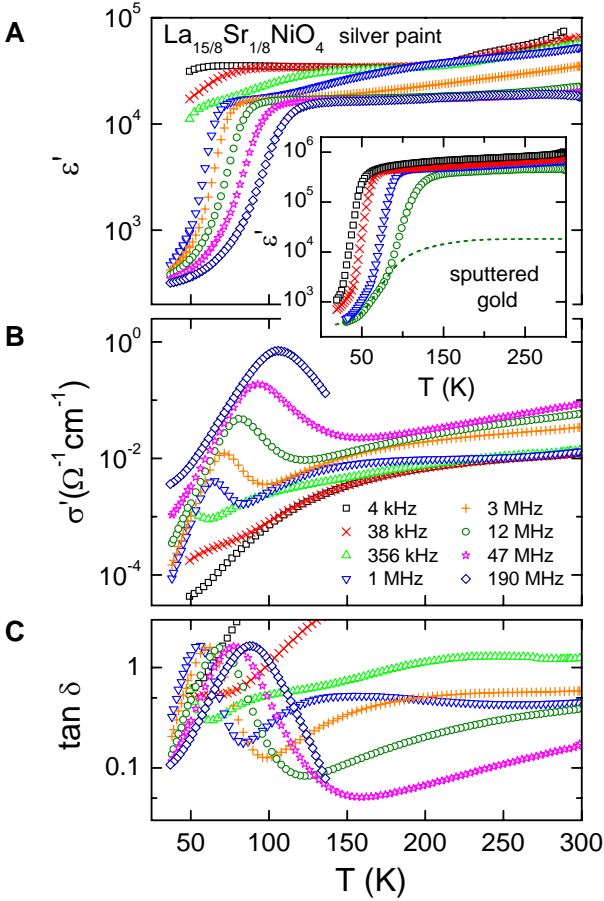

**Fig. 1** Temperature dependence of the dielectric properties of LSNO for various frequencies. The main frames show the dielectric constant (**A**), the conductivity (**B**), and the loss tangent (**C**) measured with silver-paint contacts. The inset shows the dielectric constant obtained on the same sample with sputtered gold contacts. The dashed line indicates the presence of a possible second relaxation.

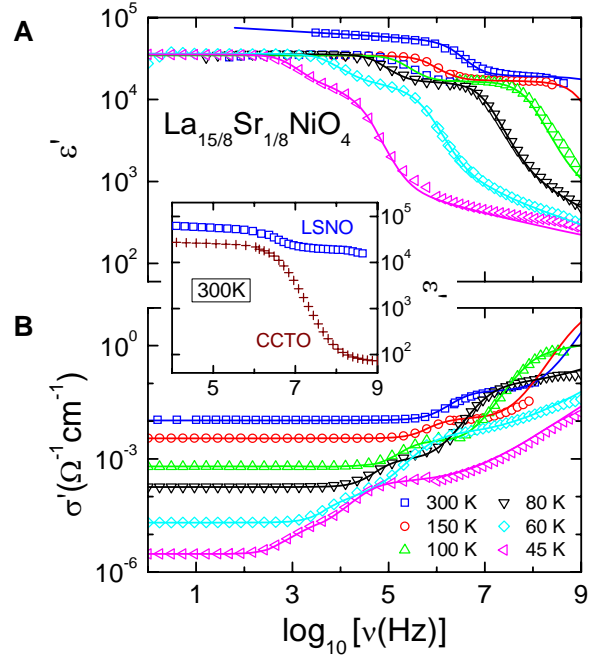

**Fig. 2** Broadband dielectric spectra of LSNO for various temperatures. (**A**): dielectric constant; (**B**): conductivity. The lines are fits with an equivalent circuit as used in refs. 8 and 12 (see text). The inset shows a comparison of $\varepsilon'(\nu)$ at room temperature in LSNO and CCTO (12).

The key result of the present work is revealed by the frequency-dependent plot of $\varepsilon'$ in Fig. 2A, extending over 8 frequency decades up to 1 GHz. Quite in contrast to CCTO (12) (see inset of Fig. 2 for a comparison), at room temperature the dielectric constant of LSNO is not affected by the main relaxation and retains its colossal magnitude up to the highest investigated frequency (430 MHz). A measurement up to higher frequencies was not possible as there these high values are out of the resolution limit of the instrument. However, already from the fact that even for 150 K the relaxation step is not shifted into the frequency window, it is clear that colossal $\varepsilon'$ values are retained even beyond 1 GHz. To obtain quantitative information, $\varepsilon'(\nu)$ and $\sigma'(\nu)$, shown in Fig. 2B, were fitted assuming an equivalent circuit just as used for CCTO (8,12). This description assumes a MW mechanism as origin of the colossal $\varepsilon'$ and the two relaxations, but also can be taken as a purely phenomenological description for parameterization purposes. It comprises two parallel RC circuits to take into account the relaxations, connected in series to the bulk element. The latter is assumed to exhibit dc conductivity and a power-law contribution, $\sigma' \propto \nu^s$, characteristic for hopping of localized charge carriers (20). Amongst other materials, such a power law with $s < 1$ is also often found in transition metal oxides (21,22) and commonly termed "universal dielectric response" (UDR) (23). The necessity to include such an ac term is best revealed by the linear increase of the 45 K curve at $\nu > 10$ MHz in the double-logarithmic plot of Fig. 2B, evidencing a sublinear power law at high frequencies. The lines shown in Fig. 2 are fits with this circuit, simultaneously performed for $\varepsilon'$ and $\sigma'$, leading to a reasonable description of the experimental data. In Fig. 3, the characteristic times of the main relaxation are shown. This Arrhenius representation reveals thermally activated behaviour of the relaxation time with an energy barrier of 67 meV. A straightforward extrapolation leads to a room temperature value of $\tau = 1.3 \times 10^{-11}$ s, corresponding to a frequency of 12 GHz. This again evidences that colossal $\varepsilon'$-values can be expected well up into the GHz regime.

It is interesting that in Fig. 2A, even at frequencies beyond the main relaxation (e.g., at 1 GHz for $\nu \leq 60$ K), with $\varepsilon' \approx 300$ the dielectric constant of LSNO still is exceptionally high, clearly exceeding the corresponding, already rather high value in CCTO of about 85 (inset of Fig. 2) (2,8,12). It should be noted that if a MW mechanism is assumed, only here the intrinsic $\varepsilon'$ is detected (10). To obtain information on the origin of this high value in LSNO, in Fig. 4 the dielectric results on $\varepsilon'(\nu)$ are shown together with those from an infrared (IR) measurement performed at room temperature. While a detailed analysis of the IR results are out of the scope of the present work, Fig. 4 clearly reveals that in LSNO, just as in CCTO (24), rather strong phonon resonances exist in the THz region pointing to a high ionic polarizability. However, these modes lead to a "static" dielectric constant (read off at the lowest frequency of the IR curve) of about 90 only. While of similar magnitude as in CCTO (24), this cannot explain the observed values of 300 at 1 GHz and 45 K in LSNO. A solution for this apparent discrepancy is provided by the UDR, which via the Kramers-Kronig relation leads to a contribution $\varepsilon'_{UDR} \propto \nu^{s-1}$, with $s < 1$, as indicated by the dash-dotted line in Fig. 4 (23). The dashed lines shown in Fig. 4 correspond to the fit curves of Fig. 2A, which were adjusted to match the IR results. Obviously, in LSNO the UDR contribution is stronger than in CCTO and plays an important role in the generation of a high $\varepsilon'$ at frequencies beyond the relaxation. When considering Fig. 4, one should be aware that the electric field directions were not identical in the dielectric (E||<110>) and IR experiment (E⊥<110>). However, within the scale of Fig. 4, only a small influence of the field direction on the strength of the phonon modes can be expected.

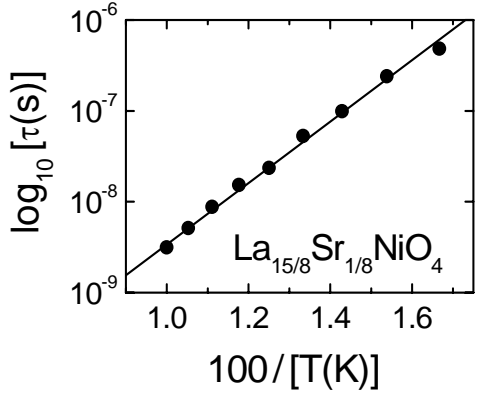

**Fig. 3** Arrhenius representation of the relaxation times. The circles show the characteristic times of the main relaxation as determined from the fits shown in Fig. 2. The line is a linear fit revealing an energy barrier of 67 meV.

Concerning the origin of the colossal $\varepsilon'$ and the observed relaxations in LSNO, we suggest the following explanations: The performed measurements with silver paint and sputtered contacts demonstrate that surface effects must play some role. Thus, part of the observed phenomena can be explained in terms of extrinsic MW effects due to contacts (Schottky diodes) or more complex surface effects (8,9,10,11,12). Grain boundaries can be excluded as in the present work single crystalline samples were investigated. However, it should be noted that in the measurements with silver-paint contacts, two relaxations are clearly detected (Figs. 1 and 2) and, thus, a second mechanism enhancing $\varepsilon'$ must be present. In addition, a close inspection of the curves obtained with sputtered gold contacts (inset of Fig. 1) reveals a shoulder at low temperatures and thus they are composed of two separate contributions, too. The first, low-temperature one may well lead to values of the order of $10^4$ (dashed line in the inset of Fig. 1) just as the main relaxation for the silver-paint electrodes (Fig. 1A). Thus, only the second, high-temperature relaxation seems to differ between silver-paint and sputtered contacts and, thus, can be assumed to be due to surface effects.

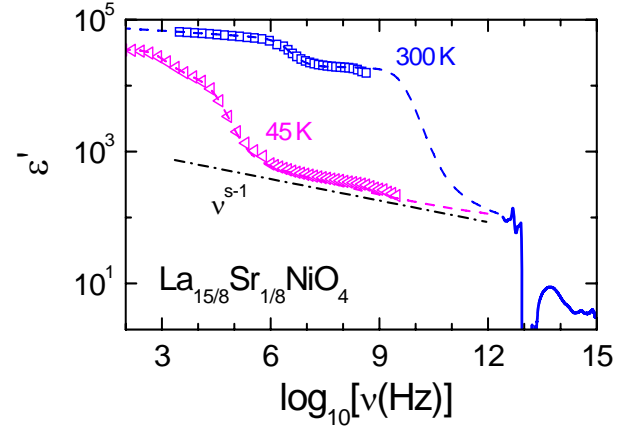

**Fig. 4** Comparison of the frequency-dependent dielectric constant of LSNO from dielectric and IR measurements. The dielectric spectra (shown as symbols) are presented for two temperatures; the IR measurements, shown as solid line were performed at room temperature. The dashed lines are the same fits as already shown in Fig. 2A, but extrapolated up to the IR range. The dash-dotted line indicates the power-law behaviour of $\varepsilon'_{UDR}$ due to hopping conductivity.

In contrast, for the first relaxation different origins should be considered: In ref. (6), a high $\varepsilon'$-magnitude and relaxational behaviour, found in La$_{5/3}$Sr$_{1/3}$NiO$_4$, was ascribed to "charge glassiness". Another very appealing possibility is heterogeneity due to spontaneously forming inhomogeneous charge distribution. Such so-called charge order, manifested as stripe-like ordering of holes, is a well established phenomenon in the system La$_{2-x}$Sr$_x$NiO$_4$ (e.g., refs. (15,16)). An enhanced dielectric constant due to charge-order-induced heterogeneity still represents a kind of MW effect. However, within this framework the origin of the MW relaxation leading to the apparently colossal $\varepsilon'$ is inherent to the material and thus this effect can be regarded as quasi-intrinsic. In addition, in this case the heterogeneity arises on a much finer scale than, e.g., for contact or grain boundary effects, which may be advantageous for application in highly miniaturized devices.

It should be noted that within the MW framework, a high intrinsic $\varepsilon'$ may be beneficial for the generation of colossal values and, thus, in CCTO it was proposed that the rather high intrinsic $\varepsilon'$, together with the small thickness of these layers leads to the apparently colossal $\varepsilon'$ (12). Within this framework, the high intrinsic $\varepsilon'$ of 300 in LSNO may favour the observation of extremely high values of $\varepsilon'$ up to the order of $10^6$ (inset of Fig. 1) in this material. The enhancement of the dielectric constant via ac conductivity, instead of increasing the ionic polarizability, thus may represent a new route to high $\varepsilon'$ values.

In summary, a detailed dielectric investigation of a charge ordered nickelate up to 1 GHz has revealed colossal values of the dielectric constant, retaining their colossal magnitude well into the technically relevant GHz frequency region. The dielectric properties of LSNO are superior to those of other colossal $\varepsilon'$ materials, including the intensely investigated CCTO. Our measurements with different contact types demonstrate that surface effects significantly contribute to the colossal $\varepsilon'$. In addition, also charge ordering is likely to play an important role in the generation of the observed unusual dielectric properties. Having in mind that, despite strong efforts, even for the much investigated CCTO no consensus has been achieved so far, it is obvious that final clarification needs further work. In any case, it is clear that here we have a material that deserves at least as much attention as CCTO, both from an application and an academic point of view.

Acknowledgements

This work was supported by the Deutsche Forschungsgemeinschaft via the Sonderforschungsbereich 484 and by the Commission of the European Communities, STREP: NUOTO, Grant No. NMP3-CT-2006-032644. Ruud Hendrikx at the Department of Materials Science and Engineering of the Delft University of Technology is acknowledged for the X-ray analysis of the sample.

Materials and Methods

The starting compositions of the samples were prepared with La:Sr:Ni cation molar ratios of 15/8 : 1/8 : 1.5. The corresponding weighted mixtures of $La_2O_3$, $SrCO_3$ and $NiO$ were reacted in a solution using $HNO_3$ as the solvent prior to the subsequent calcination and sintering processes. Since nickel has the lowest melting point, an extra 5 % of nickel was added to the starting material. The crystal was grown in a 2-mirror NEC-MHD furnace at the University of Amsterdam in an argon-oxygen (ratio 80:20) atmosphere of about 1000 mbar static pressure. Crystals resulting from this procedure were typically 3 to 4 cm long, after removing about 0.5 cm from the boule at both ends.

Crystal qualities were routinely examined by scanning electron microscopy, electron probe microanalysis and X-ray Laue backscattering, for which small amounts of sample were finely ground and analyzed with the Rietveld method. The data showed a high quality of the crystals with no traces of secondary phases. The oxygen content of the crystals selected for the present study had an almost identical stoichiometry of 4.005 – 4.010. For the crystal used for the dielectric investigation reported in this paper, the crystal orientation and lattice parameters were determined with a Panalytical X'pert Pro MRD diffractometer in a parallel beam geometry using Cu K$\alpha$1 radiation: the surface plane of the crystal on which the contacts were deposited or pasted was perpendicular to the <110> direction with one of the plane edges along the <001> direction.

For the dielectric measurements, silver paint or sputtered gold-contacts were applied at opposite sides of the plate-like samples. The dielectric properties were measured using a frequency-response analyzer (Novocontrol $\alpha$-analyzer) for frequencies $\nu < 1$ MHz and a coaxial reflection technique employing an impedance analyzer (Agilent E4991A) at $\nu > 1$ MHz (for details, see ref. (1)). In the far- and mid-infrared range, reflectivity measurements were carried out using the Bruker Fourier-transform spectrometer IFS 113v.